\documentclass[twocolumn,preprintnumbers,superscriptaddress,amsmath,amssymb,nobalancelastpage,10pt,shortbibliography,noeprint]{revtex4-1}
\usepackage{physics}
\usepackage{miller}
\usepackage{changes}
\usepackage[utf8]{inputenc}
\usepackage{multirow}
\usepackage{float}
\usepackage{bm}
\usepackage{verbatim}
\usepackage{xfrac}
\usepackage{array}
\usepackage{siunitx}
\usepackage{ulem}
\usepackage{amssymb}
\usepackage{amsmath}
\usepackage{svg}
\usepackage{hhline}

\begin{document}
	
        \title{Wavefunction tomography of topological dimer chains with long-range couplings}
        
        \author{F. Pellerin}
        \affiliation{Département de Physique, Université de Montréal, C.P. 6128, Succursale Centre-Ville, Montréal, Québec, Canada H3C 3J7}
        
        \author{R. Houvenaghel}
        \affiliation{Département de Physique, Université de Montréal, C.P. 6128, Succursale Centre-Ville, Montréal, Québec, Canada H3C 3J7}
        
        \author{W. A. Coish}
        \affiliation{Department of Physics, McGill University, 3600 rue University, Montreal, Qc H3A 2T8, Canada}
        
        \author{I. Carusotto}
        \affiliation{Pitaevskii BEC Center, INO-CNR and Dipartimento di Fisica, Università di Trento, via Sommarive 14, I-38123 Trento, Italy}
        
        \author{P. St-Jean}
        \affiliation{Département de Physique, Université de Montréal, C.P. 6128, Succursale Centre-Ville, Montréal, Québec, Canada H3C 3J7}
        
        \begin{abstract}
        The ability to tailor with a high accuracy the inter-site connectivity in a lattice is a crucial tool for realizing novel topological phases of matter. Here, we report the experimental realization of photonic dimer chains with long-range hopping terms of arbitrary strength and phase, providing a rich generalization of the celebrated Su-Schrieffer-Heeger model. Our experiment is based on a synthetic dimension scheme involving the frequency modes of an optical fiber loop platform. This setup provides direct access to both the band dispersion and the geometry of the Bloch wavefunctions throughout the entire Brillouin zone allowing us to extract the winding number for any possible configuration. Finally, we highlight a topological phase transition solely driven by a time-reversal-breaking synthetic gauge field associated with the phase of the long-range hopping, providing a route for engineering topological bands in photonic lattices belonging to the AIII symmetry class.
        \end{abstract}
        
        \maketitle
        
        \setcounter{topnumber}{2}
        \setcounter{bottomnumber}{2}
        \setcounter{totalnumber}{4}
        \renewcommand{\topfraction}{0.85}
        \renewcommand{\bottomfraction}{0.85}
        \renewcommand{\textfraction}{0.15}
        \renewcommand{\floatpagefraction}{0.7}

{\em Introduction --} Engineering materials with specific topological properties requires an acute control over the hybridization of electronic orbitals~\cite{Bradlyn2017}. As was pioneered by the Haldane model~\cite{Haldane1988}, introducing next-nearest-neighbor coupling terms with arbitrary phases strongly enriches the variety of phenomena that can be observed in topological band models~\cite{Chiu2016, Bansil2016}. Furthermore, such long-range connectivity is expected to facilitate the stabilization of strongly correlated states of matter~\cite{Kapit2010}.

The experimental implementation and control of sizable hopping terms extending beyond nearest neighbors is typically not a straightforward task~\cite{Landig2016}. In usual realizations of lattice models based on condensed matter or ultracold atomic systems, hopping typically occurs via tunneling processes mediated by
the spatial overlap of wavefunctions at different sites and is therefore dominated by short-range processes~\cite{Jaksch1998, Bloch2012}.

The situation is very different if we consider lattices extending along synthetic dimensions~\cite{Ozawa2019b}. Here, one or more of the spatial coordinates are replaced by some other internal degrees of freedom such as spin or linear momentum in ultracold atomic gases~\cite{Celi2014, Chalopin2020, An2021}, or, in photonic systems, frequency~\cite{Ozawa2016,Dutt2019}, angular momentum~\cite{Luo2017, Cardano2017a}, spatial~\cite{Lustig2019}, or temporal modes~\cite{Regensburger2011}. In the specific case of synthetic photonic lattices, a wide variety of hopping terms can be engineered through suitable modulation of the relevant degrees of freedom~\cite{Lin2018}, which has led to pioneering achievements such as the observation of the four-dimensional quantum Hall effect~\cite{Zilberberg2018}, the non-Hermitian skin effect~\cite{Weidemann2020, Xiao2020}, non-abelian excitations~\cite{Wang2021}, and treelike photonic networks~\cite{Senanian2022}.

In this Letter, we use the frequency of the photon modes in an optical fiber loop as a synthetic dimension~\cite{Dutt2019} to experimentally engineer lattices with topological bands and tunable long-range hopping terms. 
In particular, we realize one-dimensional dimerized lattices where the two sites within each unit cell are encoded in the symmetric and antisymmetric combinations of clockwise and counter-clockwise eigenmodes of a single loop. Selective hopping processes between specific pairs of sites at arbitrary distance are then introduced through a dynamical modulation of the optical fiber at a frequency resonant with their frequency difference~\cite{Dutt2020a}. This offers full control over the magnitude and the phase of the hopping terms.  

When only nearest-neighbor couplings are present, our lattice realizes the well-known Su-Schrieffer-Heeger (SSH) model~\cite{Su1979}, which displays two topologically distinct phases associated with the integer-valued winding number being $\mathcal{W}=0,1$. A wider variety of phases with $\mathcal{W}$ ranging from $-1$ to $+2$ is realized by adding $3^\mathrm{rd}$-nearest-neighbor hopping terms with specific amplitudes~\cite{Maffei2018, DErrico2020a, Vega2021}. The entire phase diagram of this Hamiltonian is reconstructed in terms of the winding number that we experimentally extract using a wavefunction tomography technique for the Bloch modes. Finally, we report the generation of a time-reversal-breaking synthetic gauge field, which allows to change the band topology without modifying the strength of the couplings---a stringent prerequisite in the conventional SSH model.

        \begin{figure*}
		\includegraphics[trim=0cm 0cm 0cm 0cm,  width=\textwidth]{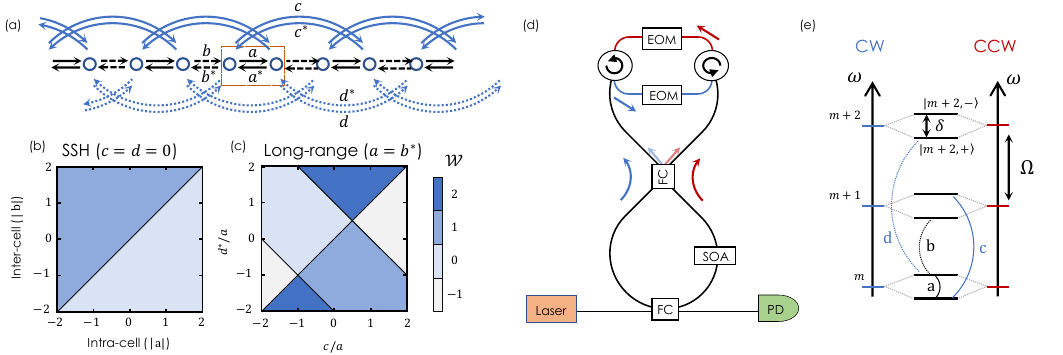}
		\caption{\textbf{Description of the experimental setup for realizing topological dimer chains in the synthetic photonic lattice.} (a) Schematic representation of a topological dimer chain. The unit cell is defined by the dotted rectangle. (b)-(c) Topological phase diagrams presenting the value of the winding number as a function of the different couplings for the SSH model (b) and the dimer chain with long-range couplings (c). (d) Experimental setup: the clockwise (CW, blue) and counter-clockwise (CCW, red) modes of an optical fiber loop are coupled with a 75:25 optical fiber coupler (FC) at the center of the loop. On top, circulators are used to separate CW and CCW modes and to modulate them independently with a pair of electro-optical phase modulators (EOMs). To optimize the quality factor of the loop, we use a semiconductor optical amplifier (SOA) to compensate the losses. The system is probed by means of an optical fiber acting as a transmission line and coupled to the main loop with a  99:1 optical fiber coupler (FC). (e) Energy spectrum of the loops in the uncoupled basis formed by a series of degenerate CW and CCW modes separated by a free spectral range $\Omega$, and in the coupled basis  of symmetric and anti-symmetric modes $\ket{m, \pm}$ separated by a frequency splitting $\delta$ and, again, a free spectral range $\Omega$.}
		\label{fig:setup}
	\end{figure*}

        \begin{figure}
            \includegraphics[trim=0cm 0cm 0cm 0cm,  width=\columnwidth]{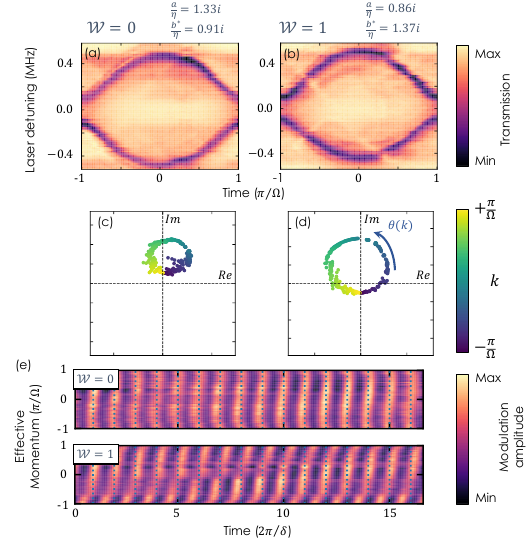}
            \caption{\textbf{Wavefunction tomography of the SSH model.} (a)-(b) Band structure measurements for the trivial (a) and topological phases (b) obtained by probing the transmitted intensity of the cavity as a function of time for different values of the laser detuning. Effective hopping terms normalized by $\eta$ are provided above each panel (in units of $\SI{}{V}$). (c)-(d) Trajectory of $g(k)$ in the complex plane for $k$ moving across the Brillouin zone in the trivial and topological cases exhibiting a winding number $\mathcal{W}=0$ and $1$, respectively. (e) Slow temporal modulation of the amplitude of the transmitted intensity peaks associated with the lower band in the trivial $\mathcal{W}=0$ (top) and topological $\mathcal{W}=1$ (bottom) cases. Vertical dashed lines indicate the periodicity of this modulation.}
            \label{fig:sshData}
        \end{figure}

{\em The topological model --} We consider in this work one-dimensional dimerized chains that present chiral symmetry, i.e. where the two atoms in each unit cell are identical and hopping processes only connect atoms in different sublattices (see Fig. \ref{fig:setup} (a)). Under the tight-binding approximation, the generic Hamiltonian describing such lattices is given by:
\begin{equation}
    H(k)=\left(\begin{array}{cc} 0 & g(k) 
    \\ g^{*}(k) & 0 \end{array} \right),
    \label{hamiltonianGraph}
\end{equation}
where the off-diagonal term $g(k)=|g(k)|e^{i\phi(k)}$ describes the hopping terms in Fourier space. Diagonalization of this Hamiltonian gives a band dispersion $E_{\pm}(k)=\pm |g(k)|$, and a phase difference $\phi(k)$ for the components of the Bloch modes on the two sublattices:
\begin{equation}
    \ket{k_{\pm}} = \frac{1}{\sqrt{2}} \begin{bmatrix} 1 \\ 
    \pm e^{-i\phi(k)}\end{bmatrix}.        
\label{eq:wavefunction}
\end{equation}
        
In this framework, the topological phases are characterized by plotting the trajectory of $g(k)$ in the complex plane throughout the Brillouin zone. The number of times the trajectory winds around the origin as $k$ spans the Brillouin zone is called the winding number $\mathcal{W}$ and is linked to the number of edge states present at the boundaries of the lattice. Although the winding number depends on the definition of the unit cell and is thus ill-defined for infinite lattices, it can be identified unambiguously for finite lattices where the definition of the unit cell is imposed by how the chain is terminated.
        
The simplest such configuration, the SSH model, has (potentially complex) nearest-neighbor hopping amplitudes only: $g(k)=a+b^{*}e^{+ikl}$ with $l$ the lattice constant. This model hosts a winding number $\mathcal{W}=0$ and $1$ for $|a|>|b|$ and $|a|<|b|$ respectively (Fig.~\ref{fig:setup} (b)). With the addition of $3^\mathrm{rd}$-nearest-neighbor hopping terms to the SSH Hamiltonian, the off-diagonal term becomes:
\begin{equation}
    g(k) = a + b^{*}e^{+ikl} + ce^{-ikl} + d^{*}e^{+2ikl}.
    \label{eq:g_of_k}
\end{equation} 

As a non-trivial consequence of these longer-range couplings, the winding number can now take values ranging from -1 to +2 upon varying the ratios between the different hopping amplitudes, see Fig.~\ref{fig:setup} (c) for the cases where $a=b^{*}$, and ratios $c/a$ and $d^{*}/a$ are real, equivalent to a model with real hopping parameters after a gauge transformation $g(k)\to e^{-i\phi_a}g(k)$ ($\phi_a=\mathrm{arg}(a)$). This choice reproduces the phase diagram from Ref.~\onlinecite{Maffei2018}. Our aim is to build a photonic platform that allows us to simulate arbitrary dimer-chain configuration, even including complex $c/a$ and $d^{*}/a$ ratios, and to extract $\mathcal{W}$ by measuring $g(k)$ throughout the Brillouin zone.

{\em The synthetic photonic lattice--} In our experiment, we use the frequency of photons confined in an optical fiber loop as a synthetic dimension. The underlying principle of this approach, inspired by Refs.~\cite{Ozawa2016,Dutt2019}, is to emulate the spatial periodicity of a lattice by exploiting the periodicity in frequency of the cavity spectrum with a period given by the free spectral range (FSR) $l=\Omega$. Coupling between specific eigenmodes is realized by locally modulating the refractive index of the cavity material with electro-optic phase modulators (EOMs) driven at a frequency equal to the corresponding mode spacing. 
        
In order to create the alternating hopping terms of a dimer chain, it is necessary to engineer a cavity with two different frequency splittings and to drive them independently. To do this, we use a single fiber loop and couple the degenerate clockwise (CW) and counter-clockwise (CCW) eigenmodes using a 75:25 optical fiber coupler (see Fig.~\ref{fig:setup} (d)). The resulting hybridized modes are symmetric and antisymmetric superpositions of the CW and CCW modes: $\ket{m, \pm} = \frac{1}{\sqrt{2}}\left(\ket{m, CCW} \pm \ket{m, CW} \right)$, where $m$ is the index of the uncoupled modes. In our setup, the splitting between $\ket{m, \pm}$ is $\delta/2\pi = 3.43~\mathrm{MHz}$ and the FSR is $\Omega /2\pi = 10.03~\mathrm{MHz}$ (Fig. \ref{fig:setup} (e)).

In order to further optimize the coupling efficiency between these eigenmodes, we use a pair of circulators that spatially separate the CW and CCW modes allowing for an independent modulation of each of them. In particular, by driving the EOMs with the same electrical signal amplitude but with a $\pi$ phase shift, $V_{cw}(t)=-V_{ccw}(t)$, we maximize the coupling between states of opposite parity while suppressing all other couplings~\cite{Dutt2020}. This choice enforces chiral symmetry by suppressing hopping processes between sites belonging to the same sublattice.

        \begin{figure*}
		\includegraphics[trim=0cm 0cm 0cm 0cm,  width=\textwidth]{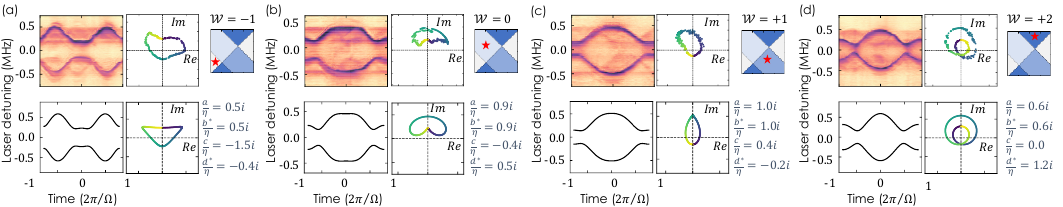}
            \caption{\textbf{Topological phases of dimer chains with long-range hopping amplitudes.} Measurements (top row) and numerical simulations (bottom row) of the band structure (left column) and of the trajectory of $g(k)$ (central column) for dimer chains with long-range hopping amplitudes. Each panel shows results for each possible value of the winding number: $\mathcal{W}=-1$ (a), $\mathcal{W}=0$ (b), $\mathcal{W}=+1$ (c) and $\mathcal{W}=+2$ (d). On the right side of each panel, we reproduce the phase diagram of Fig. \ref{fig:setup} (c), with a star indicating in which region of the phase diagram each measurement is realized, and provide the effective hopping amplitudes, normalized by $\eta$ (in units of $\SI{}{V}$).}
            \label{fig:lrData}
        \end{figure*}

{\em SSH lattices --} As a first application of our scheme, we realize an SSH Hamiltonian by driving the EOMs with a bichromatic signal $V_{ccw/cw}(t) = \pm\frac{i}{2} (V_a e^{-i\delta t} + V_b e^{-i(\Omega-\delta) t}) + \textrm{c.c.}$ where $\pm$ accounts for the $\pi$ phase shift between the two modulators on the CCW and CW paths. This modulation gives rise to effective hopping terms $a=i\eta V_{a}$ and $b^{*}=i\eta V_{b}^{*}$, where $\eta$ is related to the electro-optical constant and has units of $\SI{}{rad}~\SI{}{s}^{-1}\SI{}{V}^{-1}$~\cite{SuppMat}.

To measure the band dispersion, we probe the time-resolved transmission of the cavity using a high-bandwidth photodiode while scanning the frequency of a continuous-wave excitation laser. When the laser is in resonance with a synthetic Bloch mode $\ket{k_{\pm}}$ and the cavity decay rate $\gamma$ is much smaller than the Bloch bandwidth, the transmitted field intensity consists of a train of narrow pulses with an overall intensity modulation of frequency $\delta$:
\begin{equation}
    \frac{I_{(\pm)}(t)}{|F|^{2}} = \left[1- \frac{\kappa}{\gamma} \left(\frac{2\pi}{\Omega} \right) D_{T}(t-k)\left(1 \pm \mathrm{cos}(\delta t - \phi(k)\right) \right],\label{eq:Dirac}
\end{equation}
where $F$ is the input field amplitude, $\kappa$ is the input-output coupling strength, and $D_{T}(t)=\sum_{n}\,\delta(t-n T)$ is a Dirac comb with period $T=2\pi/\Omega$. Interestingly, in our synthetic-dimension scheme, the effective crystal momentum $k$ has units of time and corresponds to the pulse arrival time~\cite{Piccioli2022} and $T$ is the size of the effective Brillouin zone. A detailed derivation of Eq.~\eqref{eq:Dirac} and a discussion of the underlying assumptions can be found in the supplement, Ref.~\cite{SuppMat}. 

The slow modulation arises from the fact that we measure light transmitted from the CCW mode and the Bloch eigenmodes consist of linear superpositions of symmetric and anti-symmetric combinations of CW and CCW modes that oscillate at different frequencies. Hence, the signal exhibits a beating at the frequency difference $\omega_{-}- \omega_{+}=\delta$. Importantly, the phase of this modulation is exactly the relative phase of the sublattice amplitudes in the Bloch modes, allowing experimental access to the phase of the wavefunction in Eq.~\eqref{eq:wavefunction} at every $k$ point.

Figures \ref{fig:sshData} (a) and (b) show the transmitted intensity as a function of time (averaged over multiple Brillouin zones to erase the effect of the modulation) and as a function of the laser detuning for the cases $|a|>|b|$ and $|a|<|b|$, corresponding to the trivial and topological phases of the SSH model, respectively. In both cases, we clearly observe the two bands of the SSH Hamiltonian with a well-defined gap, on the order of $E_{g}\sim 200~\mathrm{kHz}$ for the parameters of the experiment.

The different topology of these two cases can be highlighted by probing the trajectory of $g(k)$ in the complex plane: $|g(k)|$ is extracted from the measurement of the band structure, and the phase $\phi(k)=\textrm{arg}(g(k))$ is obtained by Fourier transforming the slow modulation of the output signal~\cite{SuppMat}. Figure \ref{fig:sshData} (e) reports this slow modulation, as a function of time and $k$-vector, for both the trivial and topological phases. Extracting the phase of this modulation at every $k$, we can track the trajectory of $g(k)$ throughout the entire Brillouin zone (Fig.~\ref{fig:sshData} (c)-(d)), which winds around the origin in the topological case (d) but not in the trivial case (c).

 {\em Generalized SSH lattices--} Having demonstrated the ability to extract both the band structure and the Bloch wavefunctions for SSH chains, we now examine the impact of $3^\mathrm{rd}$-nearest-neighbor hopping amplitudes. In our platform, long-range hopping processes are straightforwardly implemented by adding appropriate higher-frequency components to the signal sent to the EOMs:
        \begin{equation}
        \begin{split}
            V_{ccw,cw}(t) = &\pm\frac{i}{2}(V_a e^{-i\delta t} + V_b e^{-i(\Omega-\delta) t} \\
            &+ V_c e^{-i(\Omega + \delta)t} + V_d e^{-i(2\Omega - \delta)t} ) + \textrm{c.c.},
        \end{split}
        \label{eq:voltages_LR}
        \end{equation}
where $c=- i\eta V_{c}$ and $d^{*}=-i\eta V_{d}^{*}$~\cite{SuppMat}. Here we consider cases with $a=b^{*}$ and only the ratios $c/a$ and $d^{*}/a$ are changed. Keeping these ratios real, this allows exploring the full phase diagram presented in Fig. \ref{fig:setup} (c). 

Figure \ref{fig:lrData} presents the experimental data for different ratios of long-range to nearest-neighbor hopping amplitudes. Each panel presents a specific case corresponding to one of the possible values of the winding number in the topological phase diagram: $\mathcal{W}=-1$ (panel a), $\mathcal{W}=0$ (b), $\mathcal{W}=+1$ (c) and $\mathcal{W}=+2$ (d). On the right of each panel, we indicate with a red star the position in the phase diagram of Fig. \ref{fig:setup} (c) and we give the effective hopping amplitudes used in the experiment. The experimental measurements of the band structure (top left of each panel) and of the trajectory of $g(k)$ (top center) are obtained in a similar fashion as for the previous SSH case. Finally, the bottom row presents tight-binding calculations of the band structure and the $g(k)$ trajectory using the same hopping ratios as in the experiment. We observe an excellent agreement between measurements and calculations, which validates our approach to unambiguously identify the winding number, including the full trajectory of $g(k)$.

        \begin{figure*}
        \includegraphics[trim=0cm 0cm 0cm 0cm,  width=\textwidth]{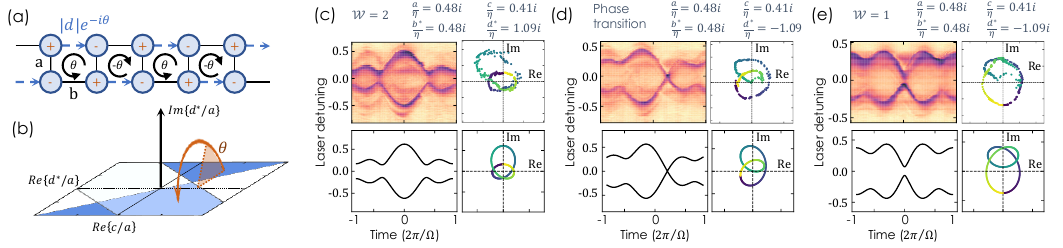}
        \caption{\textbf{Gauge-field induced topological phase transition.} (a) Schematic representation of a topological dimer chain with $a=b^{*}$, $c=0$, and complex-valued $d^{*}/a$. The two sublattices are identified with $+$ and $-$ as in Fig. \ref{fig:setup} (e). (b) Schematic representation of the trajectory in the topological phase diagram followed by the Hamiltonian as the phase of the $3^\mathrm{rd}$-nearest-neighbor hopping amplitudes ($\theta$) is scanned. (c)-(e) Measurements of the band structure (top left) and $g(k)$ trajectory (top right) as the phase of the long-range hopping is tuned from $\theta=0$ (c), to $\theta=\pi/2$ (d), and $\theta=\pi$ (e). The bottom row presents the corresponding theoretical calculation of the band structure (left) and $g(k)$ (right). For each panel, the effective hopping terms normalized by $\eta$ are provided (in units of $\SI{}{V}$).}
            \label{fig:lrPhase}
        \end{figure*}

{\em Time-reversal-breaking topological phases --} Finally, we demonstrate how the phases of the $3^\mathrm{rd}$-nearest-neighbor couplings can be tuned to break time-reversal symmetry. Adjusting these phases can induce a topological phase transition without modifying the hopping strengths. This is in sharp contrast with the conventional SSH model where it is necessary to adjust the coupling strengths to change $\mathcal{W}$, because any non-trivial phase of the hopping parameters can always be gauged away. This is a consequence of the fact that these dimer chains with broken time-reversal symmetry belong to the $AIII$ topological class rather than the $BDI$ class to which the SSH model and its extensions with time-reversal belong.

The relative phase between the different hopping terms of the Hamiltonian is experimentally realized by adding a phase to $V_{a,b,c,d}$ in Eq.~\eqref{eq:voltages_LR}. The relative phase between one of the long-range couplings and the other couplings effectively induces a synthetic gauge field for photons~\cite{Dalibard2011,Ozawa2017}. This is best seen by reformulating the dimerized chain as a ladder-like lattice (see Fig.~\ref{fig:lrPhase} (a)). In this equivalent picture, one clearly sees how the phase of $d=|d|e^{-i\theta}$ gives rise to flux plaquettes similar to those in ladder models with a staggered magnetic field. This gauge field provides an additional degree of freedom to explore a 3-dimensional topological phase space that extends the phase diagram presented in Fig.~\ref{fig:setup} (c) to the case of complex-valued $d^{*}/a$ ratios. A similar argument can of course be used for the $c$ couplings.
        
An example of a trajectory in this extended phase space is schematically depicted in Fig.~\ref{fig:lrPhase} (b), where the orange line goes through points where the magnitude of all hopping amplitudes remains constant, but the phase of $d^{*}/a$ evolves from $0$ to $\pi$. Figs.~\ref{fig:lrPhase} (c)-(e) present the measured (top) and calculated (bottom) band structures (left) and trajectories of $g(k)$ (right) along this path. Once again, except for the small deviations observed in Figs.~\ref{fig:lrPhase}(d,e) near gap closures due to the finite decay rate $\gamma$, an overall excellent agreement is found between the experiment and the theory. In particular, the breaking of time-reversal induced by the gauge field leads to a clear asymmetry of the band structure (i.e. $E(k)\neq E(-k)$). Moreover, as the gauge field is increased, we observe a closing and re-opening of the energy gap describing a topological phase transition. This transition is clearly seen through the measurement of the trajectory of $g(k)$ whose winding number changes from 2 to 1.

{\em Conclusions--} In this work, we have exploited a synthetic dimension scheme based on an optical fiber loop to realize a generalized dimer chain model including long-range and/or time-reversal breaking hopping terms. Experimental signatures of the non-trivial band topology have been obtained by reconstructing the geometry of the Bloch wavefunctions throughout the entire Brillouin zone. The natural next step will be to relate this microscopic characterization of the topology to macroscopic observables such as a driven-dissipative version of the mean chiral displacement~\cite{Villa2022, Maffei2018, Ozawa2016} and edge states in the presence of frequency-space potentials, as proposed in~\cite{Ozawa2016} and pioneered in~\cite{Dutt2022}. Setting up these tools will be instrumental in enabling investigations of more complex Hamiltonians in synthetic dimensions, involving non-Hermiticity~\cite{Gong2018, Wang2021}, higher dimensions~\cite{Cheng2023}, quantum states of light~\cite{Bello2019, Kim2021}, non-Markovian dynamics~\cite{Ricottone2020}, and/or optical nonlinearities~\cite{Ozawa2017, Pernet2022}.

{\em Note added--} During the writing of this letter, this recent work~\cite{Li2023} demonstrating the extraction of the Zak phase of SSH lattices in the synthetic frequency dimension came to our attention.

\begin{acknowledgments}
PSJ acknowledges financial support from Québec's Minstère de l'Économie, de l'Innovation et de l'Énergie. PSJ and WAC acknowledge financial support from the Fonds de Recherche--Nature et Technologies (FRQNT) and from the Natural Sciences and Engineering Research Council (NSERC). FP acknowledges financial support from FRQNT and RH from MITACS. IC acknowledges continuous collaboration with Tomoki Ozawa and Greta Villa on this topic. IC acknowledges financial support from the Provincia Autonoma di Trento and from the Q@TN initiative.    
\end{acknowledgments}

\clearpage
\onecolumngrid

\section{Supplementals: Wavefunction spectroscopy in topological dimer chains with long-range hopping}

\subsection{Description of the experimental method}
\subsubsection{Setup}

\begin{figure}[h]
    \centering
    \includegraphics[width=\textwidth]{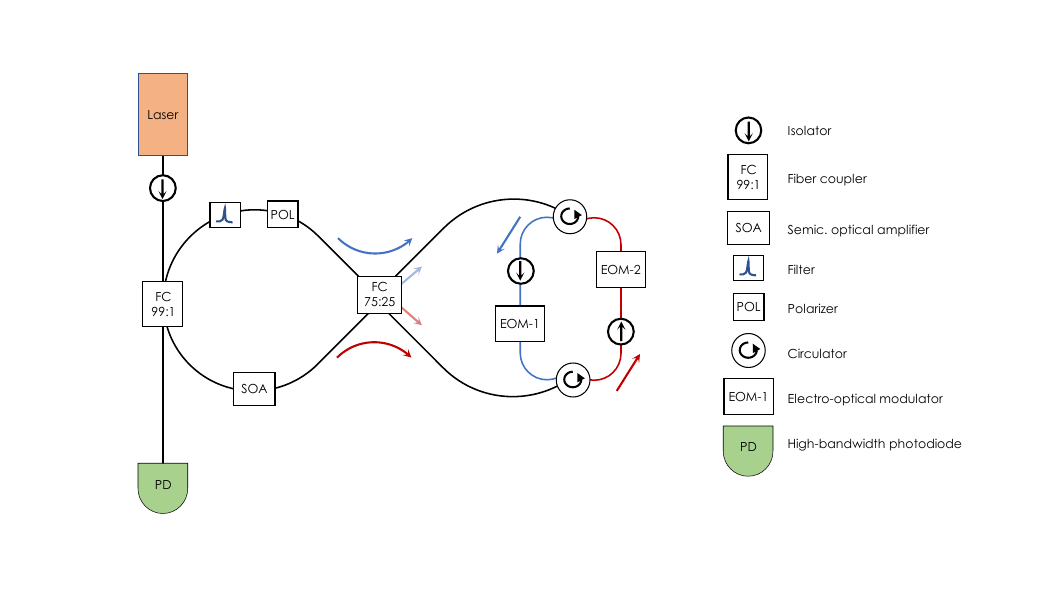}
    \caption{Schematic description of the experimental setup.}
    \label{fig:Schema}
\end{figure}

The experimental setup is shown schematically in figure \ref{fig:Schema}. All components are polarization maintaining and an in-line polarizer is added inside the loop to further ensure proper alignment of the propagating electric field polarization with the optical axis of the electro-optics phase modulators (EOMs). The cavity free spectral range (FSR) is 10.03 MHz. The staggered spectrum of the cavity, necessary to emulate dimer chains, is obtained by coupling the originally degenerate clockwise and counter-clockwise modes of the loop with a 25:75 fiber coupler. The hybridization of these modes lifts their degeneracy, giving rise to symmetric and anti-symmetric supermodes split by 3.43 MHz. Each of these pairs of symmetric and anti-symmetric supermodes forms a unit cell (i.e. a dimer) in the synthetic frequency dimension.

Since we are interested in Hamiltonians with chiral symmetry, i.e. where the hopping between sites on same sublattice vanishes, we must impose a coupling mechanism that changes the parity of the supermodes. This is achieved by spatially separating the CW and CCW propagating fields using two circulators and modulating each path independently with a dedicated EOM. The two EOMs are driven with the same electrical signal, but with a $\pi$ phase shift that ensures that symmetric supermodes only couple to anti-symmetric ones and vice-versa. Explicitly, to emulate dimer chains with long-range hoppings of amplitudes, $a$, $b$, $c$ and $d$ (as explicited in Fig. 1 (a) of the main text), the voltages sent to the two EOMs are respectively:

\begin{equation}
\begin{aligned}
&V_{ccw}(t)=\frac{i}{2}
           \left[V_{a}e^{-i\delta t}
               + V_{b}e^{-i(\Omega-\delta) t}
               + V_{c}e^{-i(\Omega+\delta)t}
               + V_{d}e^{-i(2\Omega-\delta) t}
                \right] + c.c.\\
&V_{ccw}(t)=-V_{cw}(t)
\label{eq:suppl:voltage_eom}
\end{aligned}
\end{equation}

\noindent where $V_{a,b,c,d}$ are the voltage amplitudes associated to couplings $\{a,b,c,d\}$ respectively (see subsection below for a derivation of the exact relationships). Each amplitude can be complex, with the argument $\phi_{a,b,c,d}$ describing the phase of the corresponding hopping coefficient. In our setup, the different frequency components are $f_{a}=\omega_{a}/2 \pi = 3.43$ MHz, $f_b=6.60$ MHz, $f_c=13.46$ MHz, $f_d=16.63$ MHz. 

A 1:99 input-output fiber coupler is used for realizing transmission measurements. The transmitted intensity is measured using an InGaAs photodiode with a bandwidth of \SI{10}{\giga\hertz} connected to a \SI{2}{\giga\hertz} oscilloscope. The laser is a Grade 3 Rio Orion laser with a central wavelength of \SI{1542.27}{\nano\meter} and a linewidth of \SI{3.1}{\kilo\hertz}. The frequency of the laser is modulated at \SI{40}{\hertz}, covering a range of more than \SI{30}{\mega\hertz} around the central emission frequency, so to span several FSR. The two lithium-niobate EOMs from iXblue have a bandwidth of \SI{150}{\mega\hertz} and low insertion losses. Two circulators ensure that clockwise and counter-clockwise propagating fields are dispatched to EOM 1 and 2, respectively. A semiconductor optical amplifier from Thorlabs compensates losses from the EOMs and the other components in order to achieve a high quality factor. An optical filter with a bandwith of 100 GHz is added to the loop to suppress amplification of modes far from the input laser frequency.

\subsubsection{Technique}

\begin{figure}
    \centering
    \includegraphics[width=\textwidth]{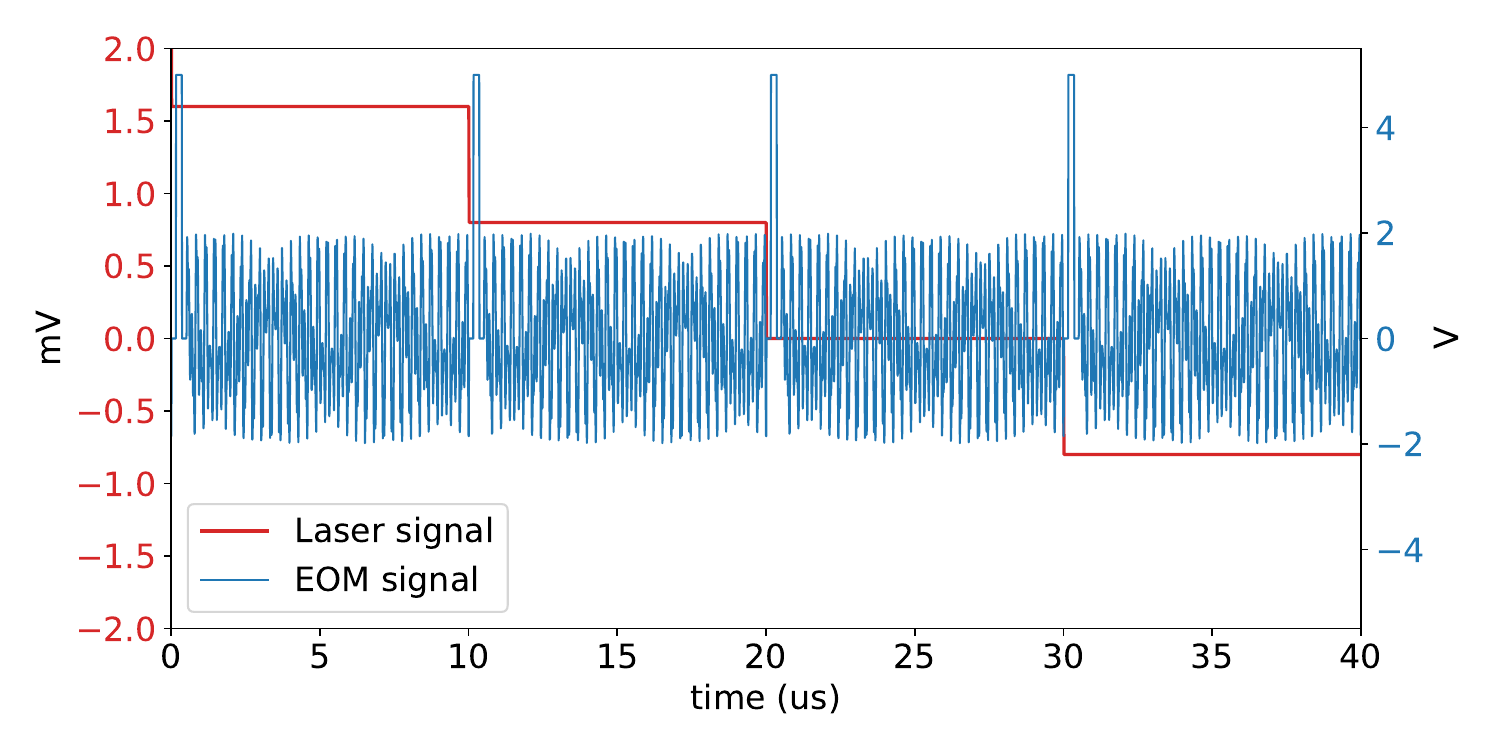}
    \caption{Example of the electrical signal (in mV) sent to the laser for adjusting its emission frequency (red line) and to the EOMs for modulating the loop (blue line).}
    \label{fig:Signaux oscilloscope}
\end{figure}

To obtain the band structure for a given set of parameters $a$, $b$, $c$, $d$, $\phi_c$, and $\phi_d$, the laser frequency is modulated by a staircase triangular waveform of frequency \SI{40}{\hertz} and amplitude \SI{0.25}{\volt}, in which each step has a duration of \SI{10}{\micro\second}. On a given step, the laser frequency is constant. The electrical signal sent to the EOMs (Eq.~\eqref{eq:suppl:voltage_eom}) is divided in bursts of \SI{10}{\micro\second} that are synchronised with the rising edge of every step (see figure \ref{fig:Signaux oscilloscope}). For each step, the signal collected by the photodiode is sent to the oscilloscope: the first half of this signal (i.e. the first $\SI{5}{\micro\second}$) is discarded to allow the system to stabilize, and the second half is sliced in fifty Brillouin zones (one Brillouin zone has a duration $1/\SI{10.03}{\mega\hertz}\approx\SI{100}{\nano\second}$). Each horizontal slice of the band structures presented in the main text is obtained by averaging the signal from these fifty Brillouin zones at every step. This averaging process ensures that the band structures do not exhibit fringes associated to the beating modulation of the Dirac comb discussed in the main text.

Furthermore, for each step, we extract the phase of $g(k)$ at every $k$-point with the following protocol. Using the band structures measure above, we determine at what laser detuning a specific time bin (i.e. a specific column of the band structure corresponding to a given $k$) presents the maximum transmission. We take the photodiode signal measured at that specific laser detuning step and keep only the data at that time bin in each Brillouin zone. This gives a slowly oscillating signal at $\SI{3.43}{\mega\hertz}$ (examples of these signals are provided in Fig. 2 (e) of the main text) that we can Fourier transform to extract the phase. This gives the phase of every point in the $g(k)$ trajectories presented in the main text.

\subsubsection{Calibration}

\paragraph{Laser frequency characterisation}

In our system, the frequency of the laser is tuned by applying a time-varying voltage to the laser. One then needs to perform a careful calibration to link the applied voltage and the output frequency of the laser. The general procedure to do this is to consider the simplest possible configuration, i.e. a single loop with only one EOM and no CW-CCW fiber coupler; the spectrum in this case is a regular cavity spectrum with equidistant modes. The FSR of this simple loop is estimated by varying the driving frequency of the EOM until the measured transmission exhibit a well-defined and symmetric band structure. For such a simple 1D lattice, the band structure must follow a cosine function and a small deviation of the driving frequency from the FSR results in an effective electric field and distorted bands (associated to Bloch oscillations). Once the driving frequency of the EOM leads to such symmetric bands, we have a precise estimation of the FSR and we can link the applied voltage to the laser frequency by doing a power-law fit on the many resonance peaks in the transmission that occur during half a period of the laser's voltage modulation. This calibration is then used to determine the frequency axis of all our band structures.

\paragraph{Quality factor}

The quality factor of the cavity is estimated with the ratio between the the resonance linewidth $\Delta\nu$ and the resonance frequency $\nu_r$. A lorentzian fit yields a FWHM of $\Delta\nu=\SI{0.075}{\mega\hertz}$. The quality factor is therefore $Q=\SI{194.4}{\tera\hertz}/\SI{0.075}{\mega\hertz}=2.6\times10^9$ and the finesse is $\mathcal{F}=133.7$.

\paragraph{EOM signals used}

We present below the amplitude and phase of the voltage signal sent to the CCW EOMs for every experiement present in this work. We use the formalism: $V_{i} = |V_{i}|e^{i \phi_{i}}$ with $i=\{a,b,c,d\}$.

\begin{center}
\begin{tabular}{ |p{3cm}|p{2cm}||p{2.5cm}|p{2.5cm}|p{2.5cm}|p{2.5cm}|  }
 \hline
 Figure   & Winding    & $|V_a|$ ($\phi_{a})$ &   $|V_b|$ ($\phi_{b})$ & $|V_c|$ ($\phi_{c})$ & $|V_d|$ ($\phi_{d})$\\ \hhline{|=|=||=|=|=|=|}
 
 \multirow{2}{8em}{Figure 2 (SSH)} & $\mathcal{W}=0$ & $\SI{1.33}{\volt}$ (0) & $\SI{0.91}{\volt}$ (0) & $\SI{0.00}{\volt}$ (0) & $\SI{0.00}{\volt}$ (0)\\
 & $\mathcal{W}=1$ & $\SI{0.86}{\volt}$ (0) & $\SI{1.37}{\volt}$ (0) & $\SI{0.00}{\volt}$ (0) & $\SI{0.00}{\volt}$ (0)\\
 \hline
 
 \multirow{4}{4em}{Figure 3 (Extended)} & $\mathcal{W}=-1$ & $\SI{0.48}{\volt}$ (0) & $\SI{0.48}{\volt}$ (0) & $\SI{1.47}{\volt}$ (0) & $\SI{0.39}{\volt}$ (0)\\
 & $\mathcal{W}=0$ & $\SI{0.86}{\volt}$ (0) & $\SI{0.86}{\volt}$ (0) & $\SI{0.37}{\volt}$ (0) & $\SI{0.52}{\volt}$ ($\pi$)\\
 & $\mathcal{W}=1$ & $\SI{0.95}{\volt}$ (0) & $\SI{0.95}{\volt}$ (0) & $\SI{0.41}{\volt}$ ($\pi$) & $\SI{0.23}{\volt}$ (0)\\
 & $\mathcal{W}=2$ & $\SI{0.57}{\volt}$ (0) & $\SI{0.57}{\volt}$ (0) & $\SI{0.00}{\volt}$ ($\pi$) & $\SI{1.24}{\volt}$ ($\pi$)\\
 \hline
 
 \multirow{3}{6em}{Figure 4 (Gauge-field)} & $\mathcal{W}=2$ & $\SI{0.48}{\volt}$ (0) & $\SI{0.48}{\volt}$ (0) & $\SI{0.41}{\volt}$ ($\pi$) & $\SI{1.09}{\volt}$ ($\pi$)\\
 & Transition & $\SI{0.48}{\volt}$ (0) & $\SI{0.48}{\volt}$ (0) & $\SI{0.41}{\volt}$ ($\pi$) & $\SI{1.09}{\volt}$ ($\pi/2$)\\
 & $\mathcal{W}=1$ & $\SI{0.47}{\volt}$ (0) & $\SI{0.47}{\volt}$ (0) & $\SI{0.41}{\volt}$ ($\pi$) & $\SI{1.09}{\volt}$ (0)\\
 \hline
\end{tabular}
    
\end{center}

\subsection{Theoretical model}

In this section we analytically derive the effective Hamiltonian of the system and its equations of motion.

\subsubsection{Derivation of the dimer chains' Hamiltonian}

Our system consists in an optical fiber loop acting as a cavity. The modes of this cavity are discrete and equally spaced by a free-spectral range $\Omega$, defined as $\frac{\Omega}{2\pi} = \frac{c}{n L}$ where $n\sim 1.45$ and $L\sim \SI{10}{\meter}$ the group index and length of our loop, and $c$ is the speed of light in vacuum. These modes can propagate either in the clockwise (CW) or counterclockwise (CCW) direction and have thus two degrees of freedom: their frequency $ \omega_m = m\Omega$ and their direction of their propagation (either CW or CCW).

Along the optical fiber the electric field is confined in one dimension, and can thus be decomposed as a linear combination of CW and CCW modes:

\begin{equation}
    \mathbf{E}(x,t)= \sum_{m} \alpha_{m} e^{-ik_mx} e^{-i\omega_m t} + \beta_m e^{ik_mx}e^{-i\omega_m t} + h.c.
\end{equation}

\noindent where $x$ is a periodic coordinate ($x=x+L$) defined along the optical fiber, $\alpha_m$ and $\beta_m$ are respectively the amplitudes of the $m$-th CW and CCW modes, and $k_m=m\Omega/c'>0$ with $c'=c/n$.

Following the usual procedure of quantization of the field, we associated $\alpha_m$ and $\alpha^{*}_m$ to bosonic operators $a_m$ and $a^{\dag}_m$, and similarly $\beta_m$ and $\beta^{*}_{m}$ to $b_m$ and $b^{\dag}_m$. This quantum notation is useful to derive the equations of motions but one should keep in mind that the modes considered in our experiment are classical coherent states.

\begin{figure}
    \centering
    \includegraphics[width=\textwidth]{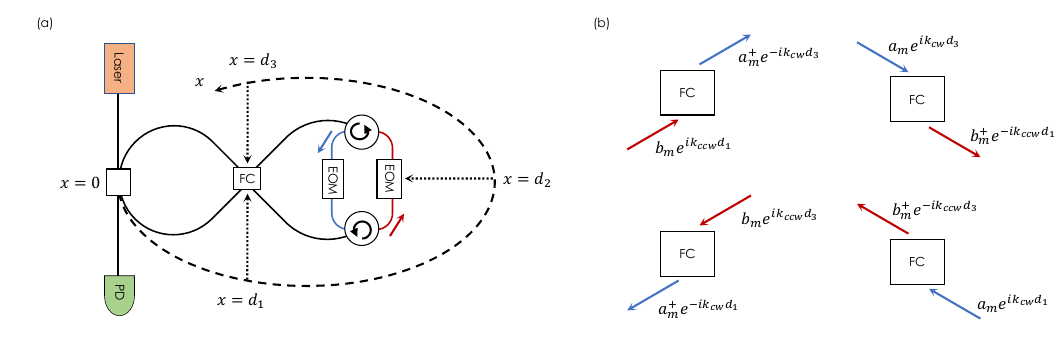}
    \caption{(a) Schematic representation describing the $x$ coordinate used for deriving the beams plitter Hamiltonian. (b) Schematic representation of the 4 terms in the beam splitter Hamiltonian in Eq.~\ref{eq:hamiltonian_bs}.}
    \label{fig:setup_coordinates}
\end{figure}

The degeneracy between the CW and CCW modes is lifted by coupling them with a 25:75 optical fiber coupler acting as a beam splitter (BS). The definition of the position coordinates $x$ along the cavity used in the derivation of this beam splitter Hamilotnian is schematically depicted in Fig.~\ref{fig:setup_coordinates} (a). The $x$ axis is periodic, oriented along the CCW direction and follows the outer fiber (i.e. the path followed by the fiber without the CW/CCW coupler). It first crosses $d_1$ at the bottom input-output ports of the CW/CCW coupler, then $d_2$ at the EOMs' position, and finally $d_3$ at the position of the top input-output ports of the CW/CCW coupler before reaching the input-output port of the cavity at $x=L$. 

The BS can destroy a photon in the CW $m$-th mode (respectively CCW $m$-th mode) entering the BS at the position $x=d_3$ and reinjects it at the position $x=d_1$ in the CCW $m$-th mode (respectively CW $m$-th mode). It is the same thing for photons entering the BS at $x=d_1$ and exiting at $x=d_3$. All 4 possible processes are schematically depicted in Fig.~\ref{fig:setup_coordinates} (b) where the red and blue arrows respectively describe the CW ($a_m$) and CCW modes ($b_m$). The momentum used in the phase factor of each modes are given by $k_{ccw} = -k_{cw} = +\frac{m\Omega}{c'}$.

The Hamiltonian describing the effect of the beam-splitter is given by:
\begin{equation}
\begin{aligned}
    H_{BS} = -\sum_{m} \Bigl[ &g \,b_{m}e^{+im\Omega\frac{d_{1}}{c'}}\, a^{\dag}_{m}e^{+im\Omega\frac{d_{3}}{c'}}
    +g^* \, a_{m}e^{-im\Omega\frac{d_{3}}{c'}}\,b^{\dag}_{m}e^{-im\Omega\frac{d_{1}}{c'}}\\
    +&g \,b_{m}e^{+im\Omega\frac{d_{3}}{c'}}a^{\dag}_{m}e^{+im\Omega\frac{d_{1}}{c'}}
    +g^ *\, a_{m}e^{-im\Omega\frac{d_{1}}{c'}}\,b^{\dag}_{m}e^{-im\Omega\frac{d_{3}}{c'}}\Bigr]
\end{aligned}
\label{eq:hamiltonian_bs}
\end{equation}
with g the coupling strength of the coupler. 
Taking into account that $e^{i\Omega L/c'}=1$ by definition of free-spectral range, we can rewrite this Hamiltonian as:
\begin{equation}
    H_{BS} = -2 \sum_{m} \left[ g^* a_{m}b^{\dag}_{m}e^{-im\Omega\frac{\Delta}{c'}}                               + g b_{m}a^{\dag}_{m}e^{+im\Omega\frac{\Delta}{c'}} \right]
\end{equation}
with $\Delta = L-d_{1}-d_{3}$ the difference between the total length of the loop $L$ and $d_{1}+d_{3}$. If the CW/CCW coupler is located symmetrically to the input-output coupler (i.e. at the same distance in the $+x$ and $-x$ directions), $\Delta=0$. However, this is not true in general, either because of thermal fluctuations or due to the difference in fiber lengths. We absorb the influence of $\Delta$ in the definition of the coupling coefficient:
\begin{equation}
        g e^{im\Omega\frac{\Delta}{c'}} = 
    \Tilde{g}.
\end{equation}
Hence, we can write $H_{BS}$ in a matrix form for the $m^{th}$ pair of modes as:
\begin{equation}
    H^{(m)}_{BS} = 
    \begin{bmatrix}
        m\Omega & -2\Tilde{g} \\
        -2\Tilde{g} & m\Omega
    \end{bmatrix}
\end{equation}
where, with no loss of generality, we have assumed for simplicity that $\tilde{g}$ is real and positive. The eigenstates are symmetric (+) and antisymmetric (-) linear combinations of the CCW and CW modes:
\begin{equation}
    \ket{m,\pm} = \frac{1}{\sqrt{2}} \left( \ket{m,CCW} \pm \ket{m, CW} \right ).
\end{equation}
With this choice for $\Tilde{g}$, the eigenfrequencies are $\omega_{m,\pm} = m\Omega \mp \frac{\delta}{2}$, with $\delta=4g$ proportional to the coupling strength (in our case $\delta/2\pi=\SI{3.43}{\mega\hertz}$), the lower (higher) frequency corresponding to the symmetric $+$ (anti-symmetric $-$) combination of CW and CCW modes. We define similarly the associated bosonic operators $c_{m,\pm}, c^{\dag}_{m, \pm}$:
\begin{equation}
    c_{m,\pm} = \frac{1}{\sqrt{2}} 
    \left[ b_{m} \pm a_{m} \right].
\end{equation}

Each of these super-modes acts as a lattice site along the synthetic frequency dimension. In order to emulate Hamiltonians with chiral symmetry, it is necessary to couple eigenmodes of opposite parity, i.e. symmetric with anti-symmetric and vice-versa. This requires modulating the CW and CCW modes independently; if they are driven with identical, in-phase electrical signals only eigenmodes of similar parity will couple. As described in the main text, this sublattice symmetry is achieved by spatially separating CW and CCW modes with circulators and modulating them independently. The Hamiltonian describing the effect of the modulators is given by:
\begin{equation}  
\begin{aligned}
    H_{EOM} &=  H_{cw} +  H_{ccw}\\
      &= 2\eta\sum_{m,n} V_{cw}(t)  a^{\dag}_{m} a_{n}e^{2i\pi(n-m) \frac{d_{2}}{L}} + 2\eta\sum_{m,n} V_{ccw}(t) b^{\dag}_{m}b_{n}e^{2i\pi(n-m) \frac{d_{2}}{L}},
\end{aligned}
\end{equation}
\noindent where $d_{2}$ is the position of the modulators along the loop (in our case, $d_{2}\sim L/2$) and $\eta$ is the electro-optical coupling coefficient which can be approximated (for small eigenmode spacing $m-n$)~\cite{Dutt2019}:
\begin{equation}
    \eta=\frac{\Omega}{4 V_{\pi}}, 
\end{equation}
where $V_{\pi} \sim \SI{2}{\volt}$ is the switching voltage amplitude of our modulators.

We impose $V_{ccw}(t) = -V_{cw}(t)=V(t)$ to minimize (maximize) coupling between states of the same (opposite) parity. This reduces the Hamiltonian to:
\begin{equation}    
    H_{EOM} = 2\eta V(t) \sum_{m,n} \left[ (-1)^{n-m} b^{\dag}_{m} b_{n} - (-1)^{n-m}  a^{\dag}_{m}a_{n} \right].
    \label{eq:hamiltonian_voltage}
\end{equation}

We now express the $H_{EOM}$ with the symmetric and anti-symmetric mode operators ($c_{m,\pm}$):
\begin{equation}    
        H_{EOM} = 2\eta V(t) \sum_{m,n}(-1)^{n-m}\left[ c^{\dag}_{m,+}c_{n,-} + c^{\dag}_{m,-}c_{n,+} \right].
\end{equation}
Note that the dependence on $\Delta$ has vanished.

The driving voltage $V(t)$ sent to the EOM has a sinusoidal form as given in the main text:
\begin{equation}
    V(t) = \frac{i}{2}
           \left[V_{a}e^{-i\delta t}
               + V_{b}e^{-i(\Omega-\delta) t}
               + V_{c}e^{-i(\Omega+\delta)t}
               + V_{d}e^{-i(2\Omega-\delta) t}
                \right] + c.c.,
\end{equation}
where $V_{a,b,c,d}$ are complex.

Injecting this expression of $V(t)$ in $H_{EOM}$ and neglecting non-resonant terms in the rotating-wave approximation (RWA), we obtain:
\begin{equation}
    \Tilde{H}_{EOM} = i\eta\sum_m\left[
      V_{a}\Tilde{c}^{\dag}_{m,-}\Tilde{c}_{m,+}
    - V_{b}\Tilde{c}^{\dag}_{m,+}\Tilde{c}_{m-1,-}
    - V_{c}\Tilde{c}^{\dag}_{m+1,-}\Tilde{c}_{m,+}
    + V_{d}\Tilde{c}^{\dag}_{m,+}\Tilde{c}_{m-2,-}
    \right] + h.c.,
    \label{eq:HEOM}
\end{equation}
where the operators are expressed in a rotating frame:
\begin{equation}
    \Tilde{c}_{m,\pm}=c_{m,\pm}e^{i\omega_{m,\pm}t}.
\end{equation}

In particular, note how in Eq.\eqref{eq:HEOM} one clearly sees a $\pi$ phase shift between the nearest-neighbor and long-range couplings that needs to be considered when defining $V(t)$.

Using the periodicity along the frequency axis, we can rewrite this Hamiltonian in reciprocal space by defining the Bloch modes:
\begin{equation}
    d_{k,\pm} = \sum_{m} e^{-ikm\Omega}\Tilde{c}_{m,\pm},
\end{equation}
where $k$ describes the crystal momentum along this synthetic dimension. In this $k$-space, $H_{EOM}$ becomes:
\begin{equation}
    \Tilde{H}_{EOM} = \sum_k H(k) = i \eta \sum_{k} d^{\dag}_{k,-}d_{k,+} \left[
    V_{a}
    + V_{b}^{*} e^{ik\Omega}
    - V_{c} e^{-ik\Omega}
    - V_{d}^{*} e^{2ik\Omega}
    \right] + h.c..
\end{equation}

This Hamiltonian can be expressed in matrix form as:
\begin{equation}
\Tilde{H}_{EOM} = \sum_k \psi_k^\dagger\cdot h(k) \cdot \psi_k
\end{equation}
\begin{equation}
h(k) = \begin{pmatrix} 0 & g(k)\\
g^*(k) & 0
\end{pmatrix};\quad 
\psi_k = \begin{pmatrix} d_{k-}\\
d_{k+}\end{pmatrix},
\end{equation}
where
\begin{equation}
    g(k)=|g(k)|e^{i\phi(k)}=i\eta\left[V_{a} + V_{b}^{*}e^{ik\Omega}
    - V_{c}e^{-ik\Omega} - V_{d}^{*}e^{2ik\Omega}
    \right].
\end{equation}
This is precisely the quantity that is plotted in the theory plots in Figs. 3 and 4 of the main text.

Direct comparison with the function $g(k)$ in the extended Hamiltonian presented in Eq. (3) of the main text, yields the following relationships between the hopping coefficients ($a,b,c,d$) and EOM voltages ($V_{a,b,c,d}$):
\begin{equation}
    \begin{aligned}
        a &= +i\eta V_{a}\\
        b^{*} &= +i\eta V_{b}^{*}\\
        c &= -i\eta V_{c}\\
        d^{*} &= -i\eta V_{d}^{*}.
    \end{aligned}
\end{equation}
Note the $\pi$ phase shift between the nearest-neighbor and long-range hopping terms, as in the main text.

These are the relationships used for relating the voltage amplitudes to effective coupling coefficients, or vice-versa, as presented in all the figures of the main text. All the voltages used in this work are provided in Table~1.
 
\subsubsection{Derivation of the electromagnetic field time evolution and time-resolved transmission}

The evolution of the electromagnetic field confined in the optical fiber loop, when driven with a CCW input field is given by the set of Langevin equations:
\begin{equation}
\begin{aligned}
    \dot{a}_{m} &= i\left[H(t),a_{m}\right]-\frac{\gamma}{2}a_{m} \\
    \dot{b}_{m} &= i\left[H(t),b_{m}\right]-\frac{\gamma}{2}b_{m} - i\sqrt{\kappa}s_{in}(t) 
\end{aligned}
\end{equation}
where $H(t) = H_{BS}+H_{EOM}(t)$, $\gamma$ is the decay rate in the loop, $\kappa$ is the input-output coupling strength and $s_{in}=Fe^{-i\omega_{L}t}$ is the input field.

Going to the coupled basis, this set of differential equations becomes, in the rotating frame:
\begin{equation}
    \dot{\Tilde{c}}_{m,\pm} = i\left[\Tilde{H}_{EOM}(t),\Tilde{c}_{m,\pm} \right] - \frac{\gamma}{2} \Tilde{c}_{m,\pm} -i \sqrt{\frac{\kappa}{2}} e^{i\omega_{m,\pm}t}s_{in}.
\end{equation}

Then, transferring to $k$-space, we have:
\begin{equation}
    \dot{d}_{k,\pm} = i\left[ H(k), d_{k,\pm} \right]- \frac{\gamma}{2} d_{k,\pm}-i \sqrt{\frac{\kappa}{2}} \sum_{m} e^{-ikm\Omega}e^{i\omega_{m,\pm}t}s_{in}.
\end{equation}

Finally, we can now go in the basis that diagonalizes $H(k)$:
\begin{equation}
    \begin{aligned}
        u_{k,\pm} = \frac{1}{\sqrt{2}}\left[ d_{k,-} \pm e^{i\phi(k)} d_{k,+}\right]
    \end{aligned}
\end{equation}
with eigenenergies $\omega_{k,\pm} = \pm|g(k)|$ and $\phi(k)$ defined as the phase of $g(k)$. In this basis, the equations of motion become:
\begin{equation}
    \dot{u}_{k,\pm} = \mp i|g(k)|u_{k,\pm} - \frac{\gamma}{2}u_{k,\pm} - i\frac{\sqrt{\kappa}}{2} s_{in} \left[ e^{+i\frac{\delta}{2}t} \pm e^{-i\frac{\delta}{2}t + i\phi(k)} \right] \sum_{m} e^{im\Omega\left(t-k\right)}
\end{equation}

Integrating this last equation over time yields:
\begin{equation}
    u_{k,\pm}(t) = u_{k,\pm}(t=0) -i\frac{\sqrt{\kappa}}{2} \int_{0}^{t}dt' e^{(\pm i|g(k)|+ \gamma/2)(t'-t)} s_{in}(t') \left[ e^{+i\frac{\delta}{2}t'} \pm e^{-i\frac{\delta}{2}t'} e^{ i\phi(k)} \right]\sum_{m} e^{im\Omega\left(t'-k\right)}
\end{equation}
where we can assume $u_{k,\pm}(t=0) = 0$.

Under the change of variable $t''=t-t'$, we can write this integral as:
\begin{equation}
    u_{k,\pm} (t) = -i\frac{\sqrt{\kappa}}{2}\left[ C_{1}^{(\pm)}(t) \pm C_{2}^{(\pm)}(t)\right]
\end{equation}
where
\begin{equation}
\begin{aligned}
    C_{1}^{(\pm)}(t) &= Fe^{-i\omega_{L}t}e^{+i\frac{\delta}{2}t} \sum_{m}\left[e^{im\Omega(t-k)} \int_{0}^{t} dt''e^{-\frac{\gamma}{2}t''} e^{i(\omega_{L} - (m\Omega +\frac{\delta}{2}\pm|g(k)|))t''}
    \right]\\
    C_{2}^{(\pm)}(t) &= Fe^{-i\omega_{L}t} e^{-i\frac{\delta}{2}t} e^{+i\phi(k)} \sum_{m}\left[e^{im\Omega(t-k)} \int_{0}^{t} dt''e^{-\frac{\gamma}{2}t''} e^{i(\omega_{L} - (m\Omega -\frac{\delta}{2}\pm|g(k)|))t''}
    \right]
    \end{aligned}
\end{equation}

These integrals are straightforwardly solved. We can assume that the upper bound to be much larger than the cavity's lifetime, i.e. $t\gg\gamma^{-1}$. In the experiment, this is realized by probing the field at times long enough to let the field reach a steady-state (in our case several microseconds after the start of each laser detuning step). This leads to:
\begin{equation}
\begin{aligned}
    C_{1}^{(\pm)}(t) &= -Fe^{-i\omega_{L}t}e^{+i\frac{\delta}{2}t} \sum_{m}\left[e^{im\Omega(t-k)} \frac{1}{i(\omega_{L} - (m\Omega +\frac{\delta}{2}\pm|g(k)|)) - \frac{\gamma}{2}}
    \right] \\
    C_{2}^{(\pm)}(t) &= -Fe^{-i\omega_{L}t}e^{-i\frac{\delta}{2}t} e^{i\phi(k)} \sum_{m}\left[e^{im\Omega(t-k)} \frac{1}{i(\omega_{L} - (m\Omega -\frac{\delta}{2}\pm|g(k)|)) - \frac{\gamma}{2}}
    \right].
\end{aligned}
\end{equation}

In this form, $C_{1}^{(\pm)}(t)$ and $C_{2}^{(\pm)}(t)$ exhibits clear resonances with a linewidth $\gamma/2$ every time the laser frequency reaches respectively $\omega_{L}=m\Omega + \delta/2 \pm|g(k)|$ and $\omega_{L}=m\Omega - \delta/2 \pm|g(k)|$. The interpretation of these periodic resonance is that the band structure of the dimer chains repeats itself along the synthetic frequency dimension at every eigenmode of the bare cavity.

Since the linewidths of the laser and of the bare modes are much smaller than $\Omega$ and $\delta$, these expressions can be further simplified. For instance, if we consider a drive frequency resonant with a band emerging from the $\Bar{m}^{th}$ symmetric mode (i.e. $\omega_{L}\simeq \Bar{m}\Omega - \delta/2$), we can simplify the equation for $u_{k,\pm}$ by neglecting completely $C_{1}^{(\pm)}$ and all terms in the summation in $C_{2}^{(\pm)}$ except the one associated to $\Bar{m}$.

\subsubsection{Input-output relations}

Finally, in order to determine the temporal profile of the transmitted field, we derive the field amplitude at the photodiode with an input-output relation where only the CCW modes are coupled to the output port:
\begin{equation}
\begin{aligned}
    s_{pd} &= s_{in} -i\sqrt{\kappa}\sum_m b_{m} \\
           &= s_{in} -i\sqrt{\frac{\kappa}{2}}\sum_m \left[\Tilde{c}_{m,+}e^{-i(m\Omega -\frac{\delta}{2})t} + \Tilde{c}_{m,-} e^{-i(m\Omega +\frac{\delta}{2})t} \right]\\
           &= s_{in} -i\sqrt{\frac{\kappa}{2}}\sum_{m,k} \left[d_{k,+} e^{-i(m\Omega -\frac{\delta}{2})t} e^{ikm\Omega} + d_{k,-} e^{-i(m\Omega +\frac{\delta}{2})t} e^{ikm\Omega} \right]\\
           &= s_{in} -i\frac{\sqrt{\kappa}}{2}\sum_{m,k} \left[(u_{k,+} - u_{k,-}) e^{-i(m\Omega -\frac{\delta}{2})t} e^{ikm\Omega} e^{-i\phi(k)} + (u_{k,+} + u_{k,-}) e^{-i(m\Omega +\frac{\delta}{2})t} e^{ikm\Omega} \right].
\end{aligned}
\end{equation}
As mentioned above, we assume the laser is in the vicinity of the $\Bar{m}^{th}$ symmetric mode so to only retain a single term in the expression for $C_1^{(\pm)}(t)$. On top of this, we assume that the laser is resonant with the $\pm$ Bloch band, so that all contributions from the other band can be neglected. Furthermore, we consider that the laser resonantly excites a single state in this band, whose wavevector $k$ is determined by a resonance condition between the laser frequency and the band dispersion $\omega_{k,\pm}$,
\begin{equation}
    \omega_L=\Bar{m}\Omega-\frac{\delta}{2}\pm|g(k)|\,.
\end{equation}
Putting all together, this gives the final expression for the field amplitude:
\begin{equation}
    u_{k,\pm}(t) = \mp i\frac{\sqrt{\kappa}}{\gamma}Fe^{-i\omega_{L}t}e^{-ik\bar{m}\Omega}e^{i(\Bar{m}\Omega -\frac{\delta}{2}) t}e^{i\phi(k)}\,
\end{equation}
Note that this approximation is only accurate when the gap $\omega_g = |\omega_{k,+} - \omega_{k,-}|$ is larger than the linewidth $\gamma$. When the gap is comparable to the linewidth, deviations appear in the extracted phase of the field and a small kink in the trajectory of $g(k)$ (as observed in Fig. 4d of the main text). 

The field at the photodiode is given by:
\begin{equation}
    s_{pd}^{(\pm)} = s_{in} \mp i\frac{\sqrt{\kappa}}{2}\sum_{m} u_{k,\pm}e^{ikm\Omega} \left[e^{-i(m\Omega -\frac{\delta}{2})t} e^{-i\phi(k)} \pm e^{-i(m\Omega +\frac{\delta}{2})t} \right].
\end{equation}
where we consider a single $k$ in resonance with the laser. We now inject $u_{k,\pm}$ obtained above and obtain:
\begin{equation}
    s_{pd}^{(\pm)} = s_{in} - \frac{\kappa}{2\gamma} Fe^{-i\omega_{L}t} \sum_{m'} e^{im' \Omega(t-k)} \left[ 1 \pm e^{-i\delta t}e^{i\phi(k)} \right]
\end{equation}
with $m'=\Bar{m} - m$. 

Then, using the identity
\begin{equation}
    D_{T}(t-a) = \frac{1}{T}\sum_m e^{i\frac{2\pi}{T}m(t-a)}
\end{equation}
where $D_{T}(t)$ is a Dirac comb of period $T$
\begin{equation}
D_T(t)=\sum_{n=-\infty}^{\infty}\delta(t-nT)\,,
\end{equation}
we obtain:
\begin{equation}
    s_{pd}^{(\pm)} = s_{in} - \frac{\kappa}{2\gamma} F \left(\frac{2\pi}{\Omega}\right)e^{-i\omega_{L}t} D_{T}(t-k)\left[ 1 \pm e^{-i\delta t} e^{i\phi(k)} \right]
\end{equation}
with $T=\frac{2\pi}{\Omega}$.

Hence the intensity measured on the photodiode for a mode $\ket{k,\pm}$ is given by:
\begin{equation}
    I_{pd}^{(\pm)} = |F|^{2} \left[1- \frac{\kappa}{\gamma} \left(\frac{2\pi}{\Omega} \right) D_{T}(t-k)\left(1 \pm \mathrm{cos}(\delta t - \phi(k)\right) \right]+\mathcal{O}(u_k^{2})
\end{equation}

This clearly demonstrates that the transmission presents a pulse train, associated to the Dirac comb with period $2\pi/\Omega$, which is modulated at frequency $2\pi/\delta$ and with a phase $\phi(k)$. Measuring the position of the Dirac comb within each single period at every laser detuning allows to extract the band structure, whereas the phase of its slow modulation across the different periods provides direct information on the relative phase of the two sublattice components of the Bloch mode wavefunctions. These features are is used in this work for extracting experimentally the Bloch wavefunction for every $k$ point and, then, the winding number. This is going to be discussed in the next Section.

For the sake of completeness, it is important to remind that this discussion assumes that a single $k$ is on resonance with the laser. If more than one $k$ are resonant with the laser (e.g. $\pm k$ in systems with time-reversal symmetry), $s_{pd}$ will consist of several pulse trains corresponding to the different $k$'s. All these pulse trains will have the same period $T$, but their temporal position will shifted in time according to their respective values of $k$. 

Note also that the shape of each pulse is exactly a $\delta$-function in the limit of a small cavity linewidth $\gamma$. In the general case, the finite $\gamma$ leads to a broadening of the pulses in time as visible in Fig.\ref{fig:sshData}(a,b).

\subsection{Experimental extraction of $\phi(k)$}

In order to obtain $\phi(k)$ from the experimental data, we perform the following procedure. For the sake of concreteness, we focus on the lower Bloch band for these calculations.
First, we determine from our experimental band structure (e.g. in Figs. 2 (a)-(b) of the main text), the laser detuning at which each $k$ mode is excited. Then, for each $k$ point we perform the following discrete Fourier transform using the intensity measured on the photodiode for the laser detuning step corresponding to the chosen $k$ value:
\begin{equation}
    \mathcal{I}(k)=\sum_n e^{i\delta t_n^{(k)}}I_{pd}(t_n^{(k)})
\end{equation}
with $t_n^{(k)}=k+nT$. In our experiment, $\delta=\SI{3.43e6}{\mega\hertz}$. The discrete Fourier transform is realized by considering only the time bins associated to the specific $k$-point considered, i.e. $t=k+nT$. The phase $\phi(k)$ is extracted by taking the argument of $\mathcal{I}(k)$:
\begin{equation}
    \phi(k)=\mathrm{arg}[\mathcal{I}(k)].
\end{equation}
The complete shape of $\phi(k)$ across the Brillouin zone is obtained by repeating this procedure for every $k$.

\end{document}